\documentclass[prb,aps,twocolumn]{revtex4}  %% revtex4 for commom submission; revtex4-2 for APS submission
\usepackage{latexsym}
\usepackage{graphicx}
\usepackage{subfigure} 
\usepackage{placeins}
\usepackage{amsmath}
\usepackage{color}
\usepackage{amsfonts}

\begin{document}

\title{ Coincidence inelastic neutron scattering for detection of two-spin magnetic correlations }

\author{Yuehua Su}
\email{suyh@ytu.edu.cn}
\affiliation{ Department of Physics, Yantai University, Yantai 264005, People's Republic of China }

\author{Shengyan Wang}
\affiliation{ Department of Physics, Yantai University, Yantai 264005, People's Republic of China }

\author{Chao Zhang}
\affiliation{ Department of Physics, Yantai University, Yantai 264005, People's Republic of China }

\begin{abstract}

Inelastic neutron scattering (INS) is one powerful technique to study the low-energy single-spin dynamics of magnetic materials. A variety of quantum magnets show novel magnetic correlations such as quantum spin liquids. These novel magnetic correlations are beyond the direct detection of INS. In this paper we propose a coincidence technique, coincidence inelastic neutron scattering (cINS), which can detect the two-spin magnetic correlations of the magnetic materials. In cINS there are two neutron sources and two neutron detectors with an additional coincidence detector. Two neutrons from the two neutron sources are incident on the target magnetic material, and they are scattered by the electron spins of the magnetic material. The two scattered neutrons are detected by the two neutron detectors in coincidence with the coincidence probability described by a two-spin Bethe-Salpeter wave function. Since the two-spin Bethe-Salpeter wave function defines the momentum-resolved dynamical wave function with two spins excited, cINS can explicitly detect the two-spin magnetic correlations of the magnetic material. Thus, it can be introduced to study the various spin valence bond states of the quantum magnets.

\end{abstract}

%\pacs{74.20.-z, 74.20.De, 74.25.-q}  % 2016-02-27

\maketitle

%%%%%%%%%%%%%%%%%%%%%%%%%%%%%%%%%%%%%%%%%%%%%%%%%%%%%%%%%%%%%%%%%%%%%%%%

\section{Introduction} \label{sec1}

The novel magnetic correlations in various quantum magnets have attracted much attention in the condensed-matter field. Quantum spin liquids with strong frustration and quantum fluctuations are one special type of example \cite{AndersonScience1987,PALeeRMP2006,BalentsNature2010,ZhouYiRMP2017}. One experimental technique in the study of these novel magnetic correlations is inelastic neutron scattering (INS), which can provide the single-spin dynamical responses of magnetic materials and thus can show the relevant physics of single-spin excitations \citep{Lovesey1984,Squires1996,FelixPrice2013,Chatterji2006,DaiRMP2015}. However, as most novel magnetic correlations in the quantum magnets are beyond that of the single-spin magnons, the spectrum of INS cannot provide explicit information on these novel magnetic correlations. It is imperative to develop experimental techniques which can explicitly detect these novel magnetic correlations.  

Recently, coincidence angle-resolved photoemission spectroscopy (cARPES) was proposed for detection of two-particle correlations of material electrons \citep{SuZhang2020}. In this paper we will follow the idea of cARPES to propose another coincidence technique, coincidence INS (cINS), which can explicitly detect the two-spin magnetic correlations of magnetic materials. There are two neutron sources and two neutron detectors in the experimental instrument of cINS, with an additional coincidence detector. The two neutron sources emit two neutrons which are incident on the target magnetic material and are scattered by the material electron spins. These two scattered neutrons are then detected by the two neutron detectors in coincidence with the coincidence probability relevant to a two-spin Bethe-Salpeter wave function. 

The two-spin Bethe-Salpeter wave function is defined as
\begin{equation}
\phi^{(ij)}_{\alpha\beta}(\mathbf{q}_1 t_1, \mathbf{q}_2 t_2) = \langle \Psi_\beta \vert T_t S^{(i)}_\perp (\mathbf{q}_1, t_1) S^{(j)}_\perp (\mathbf{q}_2, t_2) \vert \Psi_\alpha \rangle , \label{eqn1.1}
\end{equation} 
%$\phi^{(ij)}_{\alpha\beta}(\mathbf{q}_1 t_1, \mathbf{q}_2 t_2) = \langle \Psi_\beta \vert T_t S^{(i)}_\perp (\mathbf{q}_1, t_1) S^{(j)}_\perp (\mathbf{q}_2, t_2) \vert \Psi_\alpha \rangle $, 
where $\vert \Psi_\alpha \rangle$ and $\vert \Psi_\beta \rangle$ are the eigenstates of the electron spins of the target magnetic material, $S^{(i)}_\perp (\mathbf{q}, t)$ is the $i$th component of the spin operator within a perpendicular plane normal to the momentum $\mathbf{q}$, and $T_t$ is a time-ordering operator. This Bethe-Salpeter wave function describes the time dynamical evolution of the magnetic material with two spins excited at times $t_1$ and $t_2$ in time ordering. The coincidence probability of cINS can provide the Fourier transformation of the time dynamical Bethe-Salpeter wave function, with the center-of-mass frequency defined by the sum of the two transfer energies in the two-neutron scattering and the relative frequency defined by the difference of the two transfer energies. Therefore, the coincidence detection of cINS can provide the momentum-resolved dynamics of the two-spin magnetic correlations, with the physics of both the center of mass and the relative degrees of freedom of two excited spins of the magnetic material. Thus, it can be introduced to study the spin valence bond states of the quantum magnets. 

Our paper is organized as follows. In Sec. \ref{sec2} the theoretical formalism of the coincidence detection of cINS will be provided. In Sec. \ref{sec3} the coincidence probabilities of cINS for a ferromagnet and an antiferromagnet with long-range magnetic order will be presented. Discussion of the experimental detection of cINS will be given in Sec. \ref{sec4}, where a brief summary will also be provided.

%\vspace*{-0.8cm}
\section{Theoretical formalism for $\text{c}$INS} \label{sec2}

In this section we will establish the theoretical formalism for the coincidence detection of cINS. First, we will review the principle of the single-spin INS in Sec. \ref{sec2.1}. We will then provide the theoretical formalism for cINS in Sec. \ref{sec2.2}. 

\subsection{Review of INS} \label{sec2.1}

Suppose the incident neutrons have momentum $\mathbf{q}_i$ and spin $\beta_i$ with a spin distribution function $P_1(\beta_i)$.
The incident neutrons interact with the electron spins of the target magnetic material via the electron-neutron magnetic interaction 
\begin{equation}
\widehat{V}_s = \sum_{\mathbf{q}_i \mathbf{q}_f} g(\mathbf{q}) \widehat{\boldsymbol{\sigma}}_{\mathbf{q}_f \mathbf{q}_i} \cdot \mathbf{S}_{\perp}(\mathbf{q}) ,   \label{eqn2.1.1} 
\end{equation}
where $g(\mathbf{q})\equiv g F_0(\mathbf{q})$, with $g$ being an interaction constant and $F_0(\mathbf{q})$ being a magnetic form factor, and $\mathbf{q} = \mathbf{q}_f - \mathbf{q}_i$, with $\widehat{\mathbf{q}} = \frac{\mathbf{q}}{q}$. The operator $\widehat{\boldsymbol{\sigma}}_{\mathbf{q}_f \mathbf{q}_i}$ is defined for neutrons,
\begin{equation}
\widehat{\boldsymbol{\sigma}}_{\mathbf{q}_f \mathbf{q}_i} = \sum_{\beta_i \beta_f} d^{\dag}_{\mathbf{q}_f \beta_f} \boldsymbol{\sigma}_{\beta_f \beta_i} d_{\mathbf{q}_i \beta_i} , \label{eqn2.1.2} 
\end{equation}
where $d_{\mathbf{q} \beta}$ and $d^{\dag}_{\mathbf{q} \beta}$ are the respective neutron annihilation and creation operators and $\boldsymbol{\sigma}$ is the Pauli matrix. The electron spin operator $\mathbf{S}(\mathbf{q})$ is defined by
\begin{equation}
\mathbf{S}(\mathbf{q}) = \sum_{l} \mathbf{S}_l e^{-i\mathbf{q}\cdot \mathbf{R}_l}, \mathbf{S}_l =\sum_{\alpha_1 \alpha_2} c^{\dag}_{l\alpha_2} \mathbf{S}_{\alpha_2 \alpha_1} c_{l\alpha_1} , \label{eqn2.1.3}
\end{equation}
where $c_{l \alpha}$ and $c^{\dag}_{l \alpha}$ are the annihilation and creation operators of the Wannier electrons at position $\mathbf{R}_l$, respectively, and $\mathbf{S}=\frac{\boldsymbol{\sigma}}{2}$ is the spin angular momentum operator. Here we assume that the material electrons which have a dominant interaction with the incident neutrons are the local Wannier electrons. It is noted that $\mathbf{S}_{\perp}(\mathbf{q})$ is defined as    
\begin{equation}
\mathbf{S}_{\perp}(\mathbf{q}) =  \mathbf{S}(\mathbf{q}) - \widehat{\mathbf{q}} (\mathbf{S}(\mathbf{q})\cdot \widehat{\mathbf{q}}) . \label{eqn2.1.4}
\end{equation} 
A simple review of the electron-neutron magnetic interaction $\widehat{V}_s$ is given in Appendix \ref{secA1}.

One incident neutron with momentum $\mathbf{q}_i$ can be scattered by the material electrons into the state with momentum $\mathbf{q}_f$. The relevant scattering probability is defined as 
\begin{eqnarray}
\Gamma^{(1)} (\mathbf{q}_f,\mathbf{q}_i)&=&\frac{1}{Z} \sum_{\alpha\beta \beta_i\beta_f} e^{-\beta E_{\alpha}} P_1(\beta_i) \nonumber \\ 
&& \times \vert \langle \Phi_\beta \vert \widehat{S}^{(1)}(+\infty, -\infty) \vert \Phi_\alpha \rangle  \vert^2 , \label{eqn2.1.5}
\end{eqnarray}
where the initial state $\vert\Phi_\alpha\rangle=\vert \Psi_\alpha; \mathbf{q}_i \beta_i \rangle$ and the final state $\vert\Phi_\beta\rangle=\vert \Psi_\beta; \mathbf{q}_f \beta_f \rangle$ and $\vert\Psi_{\alpha}\rangle$ and $\vert\Psi_{\beta}\rangle$ are the electron eigenstates whose eigenvalues are $E_{\alpha}$ and $E_{\beta}$, respectively. $\widehat{S}^{(1)}(+\infty, -\infty)$ is the first-order expansion of the time-evolution $S$ matrix of the perturbation electron-neutron magnetic interaction $\widehat{V}_s$ and is defined as 
\begin{equation}
\widehat{S}^{(1)}(+\infty, -\infty) = -\frac{i}{\hbar} \int^{+\infty}_{-\infty} d t \widehat{V}_{I}(t) F_{\theta}(t) , \label{eqn2.1.6}
\end{equation} 
where $\widehat{V}_I(t)= e^{i H_0 t/\hbar} \widehat{V}_s e^{-i H_0 t/\hbar}$, with $H_0$ being the sum of the Hamiltonians of the material electrons and the neutrons. $F_{\theta}(t)$ defines the interaction perturbation time,
\begin{equation}
F_{\theta}(t) = \theta(t+\Delta t/2) - \theta(t-\Delta t/2) , \label{eqn2.1.7}
\end{equation}
where $\theta(t)$ is the step function.
 
It should be noted that in the above scattering probability, we have defined implicitly the initial and final states by the density matrices as follows: 
\begin{eqnarray}
&&\widehat{P}_I = \frac{1}{Z} \sum_{\alpha \beta_i}  e^{-\beta E_\alpha} P_1 (\beta_i) \vert \Psi_\alpha; \mathbf{q}_i \beta_i \rangle \langle \beta_i \mathbf{q}_i ; \Psi_\alpha \vert ,  \notag \\
&&\widehat{P}_F = \sum_{\beta \beta_f} \vert \Psi_\beta; \mathbf{q}_f \beta_f \rangle \langle \beta_f \mathbf{q}_f ; \Psi_\beta \vert . \label{eqn2.1.8}
\end{eqnarray}
In this paper we will focus on the cases where the incident neutrons are the thermal neutrons in the spin mixed state defined by 
\begin{equation}
\sum_{\beta_i} P_1 (\beta_i) \vert \beta_i \rangle \langle \beta_i \vert = \frac{1}{2} \left( \vert \uparrow \rangle \langle \uparrow \vert + \vert \downarrow \rangle \langle \downarrow \vert \right) . \label{eqn2.1.9}
\end{equation}

We introduce an imaginary-time Green's function $G(\mathbf{q},\tau) = -\sum_{ij}\langle T_\tau	S_i (\mathbf{q},\tau) S^{\dag}_j (\mathbf{q},0) \rangle (\delta_{ij} - \widehat{q}_i \widehat{q}_j )$. Its corresponding spectrum function $\chi (\mathbf{q},E)$ is defined as $\chi (\mathbf{q},E) = -2 \text{ Im} G(\mathbf{q},i\nu_n \rightarrow E + i\delta^+)$, which follows
\begin{eqnarray}
\chi (\mathbf{q},E) &=& \frac{2\pi}{Z}\sum_{\alpha\beta i j} e^{-\beta E_\alpha} \langle \Psi_\alpha \vert S^\dag_i (\mathbf{q}) \vert \Psi_\beta\rangle \langle \Psi_\beta \vert S_j (\mathbf{q}) \vert \Psi_\alpha\rangle \notag \\
&& \times (\delta_{ij} - \widehat{q}_i \widehat{q}_j ) n^{-1}_B(E) \delta(E + E_\beta - E_\alpha) . \label{eqn2.1.10}
\end{eqnarray}
The scattering probability can easily be shown to follow 
\begin{equation}
\Gamma^{(1)} (\mathbf{q}_f,\mathbf{q}_i) = \frac{\vert g(\mathbf{q}) \vert^2 \Delta t}{\hbar} \chi(\mathbf{q}, E^{(1)}) n_B (E^{(1)}) , \label{eqn2.1.11}
\end{equation}
where the transfer momentum and energy are defined as 
\begin{equation}
\mathbf{q}=\mathbf{q}_f - \mathbf{q}_i, E^{(1)} =E(\mathbf{q}_f) - E(\mathbf{q}_i), \label{eqn2.1.12}
\end{equation}
with $E(\mathbf{q}) = \frac{(\hbar q)^2}{2 m_n}$ ($m_n$ is the neutron mass), and $n_B(E)$ is the Bose distribution function. In the above derivation, we have assumed that the time interval $\Delta t$ is large and $\frac{\sin^2(a x) }{x^2} \rightarrow \pi a \delta(x)$ when $a\rightarrow +\infty$.

Let us consider the scattering cross section. We define the incident neutron flux by $J_I = n_I v_I $, where the density $n_I=\frac{1}{V_I}$ ($V_I$ is the renormalization volume for one neutron) and the velocity $v_I = \frac{\hbar q_i}{m_n}$. The scattering cross section per scatter $\sigma$ follows
\begin{equation}
J_I \sigma = \frac{1}{N_m \Delta t} \sum_{ \mathbf{q}_f} \Gamma^{(1)} (\mathbf{q}_f,\mathbf{q}_i) , \label{eqn2.1.13}
\end{equation}
where $N_m$ is the number of scatter electrons in the incident neutron beam. The double-differential scattering cross section is shown to follow
\begin{equation}
\frac{d^2\sigma}{d\Omega d E_f} = \frac{(\gamma R_e)^2}{2\pi N_m} \frac{q_f}{q_i} \vert F_0(\mathbf{q}) \vert^2 \chi(\mathbf{q}, E^{(1)}) n_B (E^{(1)}) , \label{eqn2.1.14}
\end{equation}
where $E_f$ is the energy of the scattered neutrons, $\gamma=1.91$ is a constant for the neutron gyromagnetic ratio, and $R_e$ is the classical electron radius, defined as
\begin{equation}
R_e = \frac{\mu_0 e^2}{4\pi m_e} = \frac{e^2}{4\pi \varepsilon_0 m_e c^2},  \label{eqn2.1.15}
\end{equation}
with $\mu_0$ being the free-space permeability and $\varepsilon_0$ being the vacuum permittivity. This double-differential cross section we have obtained is the same as that from Fermi's golden rule \citep{Lovesey1984,Squires1996,FelixPrice2013}. Physically, the scattering probability and the scattering cross section of INS come from the contribution of the first-order perturbation of the electron-neutron magnetic interaction.

\subsection{Theoretical formalism for $\text{c}$INS} \label{sec2.2}

\begin{figure}[ht]
\includegraphics[width=0.7\columnwidth]{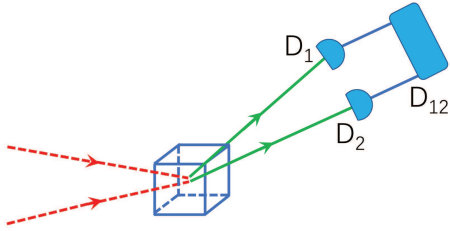} 
\caption{ (Color online) Schematic diagram of cINS. The two red dashed lines represent two incident neutrons, and the two green solid lines represent two scattered neutrons. D$_1$ and D$_2$ are two single-neutron detectors, and D$_{12}$ is a coincidence detector which records one counting when D$_1$ and D$_2$ each detect one single neutron simultaneously.}
\label{fig1}
\end{figure}

In this section we will present a coincidence technique, coincidence inelastic neutron scattering, which we call cINS. It is proposed for the detection of the two-spin magnetic correlations of the target magnetic material. The schematic diagram of cINS is shown in Fig. \ref{fig1}. There are two neutron sources which emit two neutrons with momenta $\mathbf{q}_{i_1} $ and $\mathbf{q}_{i_2}$. These two neutrons are incident on the target magnetic material and interact with the electron spins. The two incident neutrons are then scattered outside of the material into the states with momenta $\mathbf{q}_{f_1} $ and $\mathbf{q}_{f_2}$. Two single-neutron detectors detect the two scattered neutrons, and a coincidence detector records the coincidence counting probability when each of the two single-neutron detectors detects one single neutron simultaneously.  

The coincidence counting probability of the two scattered neutrons is described by 
\begin{eqnarray}
\hspace*{-0.5cm}\Gamma^{(2)}(\mathbf{q}_{f_1}\mathbf{q}_{f_2}, \mathbf{q}_{i_1}\mathbf{q}_{i_2}) && = \frac{1}{Z} \sum_{\alpha\beta \beta_i\beta_f} e^{-\beta E_{\alpha}} P_2(\beta_{i_1},\beta_{i_2}) \nonumber \\
&& \times \vert \langle \Phi_\beta \vert \widehat{S}^{(2)}(+\infty, -\infty) \vert \Phi_\alpha \rangle  \vert^2 , \label{eqn2.2.1}
\end{eqnarray}
where the initial state $\vert\Phi_\alpha\rangle=\vert \Psi_\alpha; \mathbf{q}_{i_1} \beta_{i_1} \mathbf{q}_{i_2} \beta_{i_2} \rangle$ and the final state $\vert\Phi_\beta\rangle=\vert \Psi_\beta; \mathbf{q}_{f_1} \beta_{f_1} \mathbf{q}_{f_2}\beta_{f_2} \rangle$. $P_2(\beta_{i_1},\beta_{i_2})$ defines the spin distribution function of the incident thermal neutrons. In the following, we will consider the cases with $P_2(\beta_{i_1},\beta_{i_2}) = P_1(\beta_{i_1}) P_1(\beta_{i_2})$. $\widehat{S}^{(2)}(+\infty, -\infty)$ is the second-order expansion of the time-evolution $S$ matrix and is defined by 
\begin{eqnarray}
&& \widehat{S}^{(2)}(+\infty, -\infty) \nonumber \\
&& = \frac{1}{2!}\left(-\frac{i}{\hbar}\right)^2 \iint^{+\infty}_{-\infty} d t_1 d t_2 T_t [\widehat{V}_{I}(t_1)\widehat{V}_{I}(t_2) ] F_{\theta}(t_1,t_2) . \nonumber \\
&&  \label{eqn2.2.2}
\end{eqnarray}
Here the time function $F_{\theta}(t_1,t_2)$ is defined as $F_{\theta}(t_1,t_2) = F_{\theta}(t_1) F_{\theta}(t_2)$. Physically, the coincidence probability of cINS is determined by the second-order perturbation of the electron-neutron magnetic interaction.
 
Following the theoretical treatment for cARPES \citep{SuZhang2020}, we introduce the two-spin Bethe-Salpeter wave function defined in Eq. (\ref{eqn1.1}).  
With the two-spin Bethe-Salpeter wave function, we can show that the coincidence probability of cINS can be expressed as 
\begin{equation}
\Gamma^{(2)} = \Gamma^{(2)}_1 + \Gamma^{(2)}_2 , \label{eqn2.2.4}
\end{equation}
where
\begin{eqnarray*}
\Gamma^{(2)}_1 &=& \frac{1}{Z}\sum_{\alpha\beta \beta_i\beta_f} e^{-\beta E_\alpha} P_1(\beta_{i_1}) P_1(\beta_{i_2})  \\
&& \times \frac{1}{\hbar^4} \bigg\vert \iint^{+\infty}_{-\infty} d t_1 d t_2 M_{\alpha\beta,1}(t_1,t_2) F_{\theta}(t_1,t_2)  \bigg\vert^2 , \\
\Gamma^{(2)}_2 &=& \frac{1}{Z}\sum_{\alpha\beta \beta_i\beta_f} e^{-\beta E_\alpha} P_1(\beta_{i_1}) P_1(\beta_{i_2}) \\
&& \times \frac{1}{\hbar^4} \bigg\vert \iint^{+\infty}_{-\infty} d t_1 d t_2 M_{\alpha\beta,2}(t_1,t_2) F_{\theta}(t_1,t_2)  \bigg\vert^2 .
\end{eqnarray*}
Here the matrix elements $M_{\alpha\beta,1}$ and $M_{\alpha\beta,2}$ are defined as 
\begin{eqnarray*}
M_{\alpha\beta,1} &=& g(\mathbf{q}_1) g(\mathbf{q}_2) \sum_{ij} \phi^{(ij)}_{\alpha\beta}(\mathbf{q}_1 t_1, \mathbf{q}_2 t_2) \\
&& \times \sigma^{(i)}_{\beta_{f_1}\beta_{i_1}} \sigma^{(j)}_{\beta_{f_2}\beta_{i_2}} e^{i(E^{(2)}_1 t_1 +E^{(2)}_2 t_2)/\hbar } ,   \\
M_{\alpha\beta,2} &=& g(\overline{\mathbf{q}}_1) g(\overline{\mathbf{q}}_2) \sum_{ij} \phi^{(ij)}_{\alpha\beta}(\overline{\mathbf{q}}_1 t_1, \overline{\mathbf{q}}_2 t_2) \\
&& \times \sigma^{(i)}_{\beta_{f_1}\beta_{i_2}} \sigma^{(j)}_{\beta_{f_2}\beta_{i_1}} e^{i(\overline{E}^{(2)}_1 t_1 +\overline{E}^{(2)}_2 t_2)/\hbar } , 
\end{eqnarray*}
where the transfer momenta are defined as 
\begin{eqnarray}
&& \mathbf{q}_1 = \mathbf{q}_{f_1}-\mathbf{q}_{i_1} , \mathbf{q}_2 = \mathbf{q}_{f_2}-\mathbf{q}_{i_2}, \nonumber \\
&& \overline{\mathbf{q}}_1 = \mathbf{q}_{f_1}-\mathbf{q}_{i_2} , \overline{\mathbf{q}}_2 = \mathbf{q}_{f_2}-\mathbf{q}_{i_1} , \label{eqn2.2.5}
\end{eqnarray}
and the transfer energies are defined as
\begin{eqnarray}
\hspace*{-0.5cm} E^{(2)}_1 = E(\mathbf{q}_{f_1}) - E(\mathbf{q}_{i_1}) , E^{(2)}_2 = E(\mathbf{q}_{f_2}) - E(\mathbf{q}_{i_2}) , &&  \nonumber \\
\hspace*{-0.5cm} \overline{E}^{(2)}_1 = E(\mathbf{q}_{f_1}) - E(\mathbf{q}_{i_2}) ,  \overline{E}^{(2)}_2 = E(\mathbf{q}_{f_2}) - E(\mathbf{q}_{i_1}) .   \label{eqn2.2.6}  
\end{eqnarray}
Physically, there are two different classes of microscopic neutron scattering processes involved in the coincidence scattering. One is with the state changes of the two neutrons as $\vert \mathbf{q}_{i_1}\beta_{i_1}\rangle \rightarrow \vert \mathbf{q}_{f_1}\beta_{f_1}\rangle$ and $\vert \mathbf{q}_{i_2}\beta_{i_2}\rangle \rightarrow \vert \mathbf{q}_{f_2}\beta_{f_2}\rangle$, and the other one is with $\vert \mathbf{q}_{i_1}\beta_{i_1}\rangle \rightarrow \vert \mathbf{q}_{f_2}\beta_{f_2}\rangle$ and $ \vert \mathbf{q}_{i_2}\beta_{i_2}\rangle \rightarrow \vert \mathbf{q}_{f_1}\beta_{f_1}\rangle$. 
The matrix elements $M_{\alpha\beta,1}$ and $M_{\alpha\beta,2}$ and the corresponding coincidence probabilities $\Gamma^{(2)}_1$ and $\Gamma^{(2)}_2$ describe these two different classes of microscopic neutron scattering processes, respectively. It should be noted that here we have ignored the quantum interference of these two different scattering contributions as they come from different scattering channels of energy transfer with energy-conservation-like resonance features at different energies.

We define the center-of-mass time $t_c = \frac{1}{2}(t_1 + t_2)$ and the relative time $t_r = t_1 - t_2$ and denote the two-spin Bethe-Salpeter wave function $\phi^{(ij)}_{\alpha\beta}(\mathbf{q}_1, \mathbf{q}_2; t_c, t_r) = \phi^{(ij)}_{\alpha\beta}(\mathbf{q}_1 t_1, \mathbf{q}_2 t_2)$. We can introduce the Fourier transformations of $\phi^{(ij)}_{\alpha\beta}(\mathbf{q}_1, \mathbf{q}_2; t_c, t_r)$ as follows: 
\begin{eqnarray*}
&& \phi^{(ij)}_{\alpha\beta}(\mathbf{q}_1, \mathbf{q}_2; t_c, t_r) \\
&& = \iint^{+\infty}_{-\infty} \frac{d\Omega d\omega}{(2\pi)^2} \phi^{(ij)}_{\alpha\beta}(\mathbf{q}_1, \mathbf{q}_2; \Omega, \omega) e^{-i\Omega t_c - i\omega t_r} , \\
&& \phi^{(ij)}_{\alpha\beta}(\mathbf{q}_1, \mathbf{q}_2; \Omega, \omega) \\
&& = \iint^{+\infty}_{-\infty} d t_c d t_r \phi^{(ij)}_{\alpha\beta}(\mathbf{q}_1, \mathbf{q}_2; t_c, t_r) e^{i\Omega t_c + i\omega t_r} .
\end{eqnarray*}
For the incident thermal neutrons in the spin mixed state defined by Eq. (\ref{eqn2.1.9}), the coincidence probability is shown to follow
\begin{eqnarray}
\hspace*{-0.5cm}\Gamma^{(2)} &&= \frac{1}{\hbar^4}\frac{1}{Z}\sum_{\alpha\beta i j} e^{-\beta E_{\alpha}} \nonumber \\
&& \times [ C_1 \big\vert \phi^{(ij)}_{\alpha\beta,1}(\mathbf{q}_1 , \mathbf{q}_2)  \big\vert^2 + C_2 \big\vert \phi^{(ij)}_{\alpha\beta,2}(\overline{\mathbf{q}}_1 , \overline{\mathbf{q}}_2) \big\vert^2 ]  ,   \label{eqn2.2.7}
\end{eqnarray}
where the two factors are defined as 
\begin{equation}
C_1 = \vert g(\mathbf{q}_1) g(\mathbf{q}_2) \vert^2, C_2 = \vert g(\overline{\mathbf{q}}_1) g(\overline{\mathbf{q}}_2) \vert^2 , \label{eqn2.2.8}
\end{equation}
and the two wave functions $\phi^{(ij)}_{\alpha\beta,1}(\mathbf{q}_1 , \mathbf{q}_2)$ and $\phi^{(ij)}_{\alpha\beta,2}(\overline{\mathbf{q}}_1 , \overline{\mathbf{q}}_2)$ are defined as 
\begin{eqnarray}
&&\phi^{(ij)}_{\alpha\beta,1}(\mathbf{q}_1 , \mathbf{q}_2) \nonumber \\
 && = \iint^{+\infty}_{-\infty} \frac{d\Omega d\omega}{(2\pi)^2} \phi^{(ij)}_{\alpha\beta}(\mathbf{q}_1 , \mathbf{q}_2; \Omega,\omega) Y_1 (\Omega, \omega) ,  \label{eqn2.2.9} \\
&&\phi^{(ij)}_{\alpha\beta,2}(\overline{\mathbf{q}}_1 , \overline{\mathbf{q}}_2) \nonumber \\
&&  = \iint^{+\infty}_{-\infty} \frac{d\Omega d\omega}{(2\pi)^2} \phi^{(ij)}_{\alpha\beta}(\overline{\mathbf{q}}_1 , \overline{\mathbf{q}}_2; \Omega,\omega) Y_2 (\Omega, \omega) . \label{eqn2.2.10}
\end{eqnarray} 
Here the functions $Y_1 (\Omega, \omega)$ and $Y_2 (\Omega, \omega)$ are given by 
\begin{eqnarray}
Y_1 (\Omega, \omega) &=& \frac{\sin[(E^{(2)}_1/\hbar -\Omega/2 -\omega ) \Delta t/2]}{(E^{(2)}_1/\hbar -\Omega/2 -\omega )/2} \nonumber \\
&& \times \frac{\sin[(E^{(2)}_2/\hbar -\Omega/2 +\omega ) \Delta t/2]}{(E^{(2)}_2/\hbar -\Omega/2 +\omega )/2} , \label{eqnA2.2.11} \\ 
Y_2 (\Omega, \omega) &=& \frac{\sin[(\overline{E}^{(2)}_1/\hbar -\Omega/2 -\omega ) \Delta t/2]}{(\overline{E}^{(2)}_1/\hbar -\Omega/2 -\omega )/2} \nonumber \\
&& \times \frac{\sin[(\overline{E}^{(2)}_2/\hbar -\Omega/2 +\omega ) \Delta t/2]}{(\overline{E}^{(2)}_2/\hbar -\Omega/2 +\omega )/2} . \label{eqn2.2.12}
\end{eqnarray} 
In large, but finite, $\Delta t$, we can make the approximation that $\int_{-\Delta t/2}^{\Delta t/2}d t_2 \int_{-\Delta t/2}^{\Delta t/2}d t_1 \rightarrow \int_{-\Delta t/2}^{\Delta t/2}d t_c \int_{-\Delta t/2}^{\Delta t/2}d t_r $. In this case the functions $Y_1 (\Omega, \omega)$ and $Y_2 (\Omega, \omega)$ can be approximated as 
\begin{eqnarray}
\hspace*{-0.5cm} Y_1 (\Omega, \omega) &=& \frac{\sin[(\Omega - E^{(2)}_1/\hbar - E^{(2)}_2/\hbar ) \Delta t/2]}{(\Omega - E^{(2)}_1/\hbar - E^{(2)}_2/\hbar )/2} \nonumber \\
&& \times \frac{\sin[(\omega - E^{(2)}_1/2\hbar + E^{(2)}_2/2\hbar )\Delta t/2]}{(\omega - E^{(2)}_1/2\hbar + E^{(2)}_2/2\hbar )/2} , \label{eqn2.2.13} \\ 
\hspace*{-0.5cm} Y_2 (\Omega, \omega) &=& \frac{\sin[(\Omega - \overline{E}^{(2)}_1/\hbar - \overline{E}^{(2)}_2/\hbar ) \Delta t/2]}{(\Omega - \overline{E}^{(2)}_1/\hbar - \overline{E}^{(2)}_2/\hbar )/2} \nonumber \\
&& \times \frac{\sin[(\omega - \overline{E}^{(2)}_1/2\hbar + \overline{E}^{(2)}_2/2\hbar )\Delta t/2]}{(\omega - \overline{E}^{(2)}_1/2\hbar + \overline{E}^{(2)}_2/2\hbar )/2}. \label{eqn2.2.14}
\end{eqnarray} 
 
\begin{widetext}
In the limit with $\Delta t \rightarrow +\infty$, it can be shown that 
\begin{equation}
\Gamma^{(2)} = \frac{1}{\hbar^4}\frac{1}{Z}\sum_{\alpha\beta i j} e^{-\beta E_{\alpha}} [ C_1 \big\vert \phi^{(ij)}_{\alpha\beta}(\mathbf{q}_1 , \mathbf{q}_2; \Omega_1,\omega_1)  \big\vert^2 + C_2 \big\vert \phi^{(ij)}_{\alpha\beta}(\overline{\mathbf{q}}_1 , \overline{\mathbf{q}}_2; \Omega_2, \omega_2) \big\vert^2 ]  , \label{eqn2.2.15} 
\end{equation}
where the transfer frequencies are defined as
\begin{equation}
\Omega_1 = \frac{1}{\hbar}(E^{(2)}_1 + E^{(2)}_2 ), \omega_1 = \frac{1}{2\hbar}(E^{(2)}_1 - E^{(2)}_2 ),  \Omega_2 = \frac{1}{\hbar}(\overline{E}^{(2)}_1 + \overline{E}^{(2)}_2 ), \omega_2 = \frac{1}{2\hbar}(\overline{E}^{(2)}_1 - \overline{E}^{(2)}_2 ) . \label{eqn2.2.16}
\end{equation}
The coincidence probability $\Gamma^{(2)}$ in Eq. (\ref{eqn2.2.15}) shows that cINS can explicitly detect the frequency Bethe-Salpeter wave function, which describes the dynamical magnetic physics of the target material with two-spin excitations involved. This can be seen more clearly from the following spectrum expression of the frequency Bethe-Salpeter wave function: 
\begin{equation} 
\phi^{(ij)}_{\alpha\beta}\left(\mathbf{q}_1, \mathbf{q}_2; \Omega, \omega \right)  = 2\pi \delta \left[\Omega + \left( E_{\beta} - E_{\alpha}\right)/\hbar \right] \phi^{(ij)}_{\alpha\beta}\left(\mathbf{q}_1, \mathbf{q}_2; \omega \right) , \label{eqn2.2.17} 
\end{equation} 
where $\phi^{(ij)}_{\alpha\beta}\left(\mathbf{q}_1, \mathbf{q}_2; \omega \right)$ follows
\begin{equation}
\phi^{(ij)}_{\alpha\beta}\left(\mathbf{q}_1, \mathbf{q}_2; \omega \right) =\sum_{\gamma} \left[ \frac{ i \langle\Psi_{\beta} \vert S^{(i)}_\perp (\mathbf{q}_1)  \vert \Psi_{\gamma} \rangle \langle \Psi_{\gamma} \vert  S^{(j)}_\perp (\mathbf{q}_2) \vert \Psi_{\alpha} \rangle} {\omega + i\delta^+ + (E_{\alpha} + E_{\beta} - 2 E_{\gamma} )/2\hbar} - \frac{ i \langle\Psi_{\beta} \vert S^{(j)}_\perp (\mathbf{q}_2) \vert \Psi_{\gamma} \rangle \langle \Psi_{\gamma} \vert  S^{(i)}_\perp (\mathbf{q}_1) \vert \Psi_{\alpha} \rangle} {\omega - i\delta^+ - (E_{\alpha} + E_{\beta} - 2 E_{\gamma} )/2\hbar} \right] . \label{eqn2.2.18}
\end{equation}
\end{widetext}
Obviously, the frequency Bethe-Salpeter wave function involves the following dynamical magnetic physics of two spins of the target magnetic material: (1) the center-of-mass dynamics of the two spins described by $\delta \left[\Omega + \left( E_{\beta} - E_{\alpha}\right)/\hbar \right]$, which shows the transfer energy conservation with the center-of-mass degrees of freedom involved, and (2) the two-spin relative dynamics $\phi^{(ij)}_{\alpha\beta}\left(\mathbf{q}_1, \mathbf{q}_2; \omega \right)$, which has resonance structures peaked at $\mp ( E_\alpha + E_\beta - 2 E_\gamma )/2\hbar $ with weights $\langle\Psi_{\beta} \vert S^{(i)}_\perp (\mathbf{q}_1)  \vert \Psi_{\gamma} \rangle \langle \Psi_{\gamma} \vert  S^{(j)}_\perp (\mathbf{q}_2) \vert \Psi_{\alpha} \rangle$ and $\langle\Psi_{\beta} \vert S^{(j)}_\perp (\mathbf{q}_2) \vert \Psi_{\gamma} \rangle \langle \Psi_{\gamma} \vert  S^{(i)}_\perp (\mathbf{q}_1) \vert \Psi_{\alpha} \rangle$, respectively. Therefore, cINS can provide the momentum-resolved dynamical two-spin magnetic correlations of the target magnetic material.

\section{Coincidence probabilities of the ferromagnet and antiferromagnet} \label{sec3} 

In this section we will study the coincidence probabilities of cINS for a ferromagnet and an antiferromagnet which have long-range magnetic order with well-defined magnon excitations. 

Provided that (1) the two incident neutrons are independent following a spin distribution function as $P_2 (\beta_{i_1},\beta_{i_2}) = P_1(\beta_{i_1}) P_1(\beta_{i_2})$ and (2) the single-spin magnetic excitations of the target material have well-defined momenta and are decoupled from each other, the coincidence probability of cINS has a simple product behavior, which can be expressed mathematically as 
\begin{eqnarray}
\Gamma^{(2)} &=& \Gamma^{(1)} (\mathbf{q}_{f_1},\mathbf{q}_{i_1}) \cdot \Gamma^{(1)} (\mathbf{q}_{f_2},\mathbf{q}_{i_2}) \notag \\
&+& \Gamma^{(1)} (\mathbf{q}_{f_1},\mathbf{q}_{i_2}) \cdot \Gamma^{(1)} (\mathbf{q}_{f_2},\mathbf{q}_{i_1}) . \label{eqn3.1}
\end{eqnarray}
This is a general result which can be exactly proven from the definitions of the scattering probability of INS and the coincidence probability of cINS, Eq. (\ref{eqn2.1.5}) and (\ref{eqn2.2.1}).

We will consider localized spin magnetic systems with a cubic crystal lattice, the Hamiltonians of which are defined by 
\begin{equation}
H = \frac{J}{2}\sum_{l\delta} \mathbf{S}_l \cdot \mathbf{S}_{l+\delta} , \label{eqn3.2}
\end{equation} 
where $\boldsymbol{\delta} = \pm a \mathbf{e}_x, \pm a \mathbf{e}_y, \pm a \mathbf{e}_z$. The localized spins are in a low-temperature ordering state with the magnetic moments ordered along the $\mathbf{e}_z$ axis.

\subsection{Ferromagnet} \label{sec3.1}

Let us consider a ferromagnet with $J<0$. We introduce the Holstein-Primakoff transformation, 
$S_l^+ = \sqrt{2S - a_l^\dag a_l} a_l, S_l^- = a_l^\dag \sqrt{2S - a_l^\dag a_l}, S_l^z = S - a_l^\dag a_l $, where $a_l$ and $a^{\dag}_l$ are the bosonic magnon operators. In linear spin-wave theory, the spin Hamiltonian can be approximated as 
\begin{equation}
H_{FM} = \sum_{\mathbf{k}} \varepsilon_\mathbf{k} a_\mathbf{k}^\dag a_\mathbf{k} , \label{eqn3.1.1}
\end{equation}
where $\varepsilon_{\mathbf{k}}= \vert J \vert zS(1-\gamma_{\mathbf{k}})$, with $\gamma_{\mathbf{k}}=\frac{1}{z}\sum_{\boldsymbol{\delta}} e^{i\mathbf{k}\cdot \boldsymbol{\delta}}$ and  coordination number $z=6$.  Here $a_\mathbf{k} =\frac{1}{\sqrt{N}}\sum_l a_l  e^{-i\mathbf{k}\cdot \mathbf{R}_l}$.

Let us first study the scattering probability of the single-spin INS. Suppose the incident thermal neutrons are in the spin mixed state defined by Eq. (\ref{eqn2.1.9}). It can be shown from Eq. (\ref{eqn2.1.11}) that the scattering probability $\Gamma^{(1)}$ follows 
\begin{eqnarray}
 && \Gamma^{(1)}_{FM} (\mathbf{q}_f,\mathbf{q}_i) \notag \\
= && \frac{\vert g(\mathbf{q}) \vert^2 \Delta t }{\hbar}  n_B (E^{(1)}) \left[\chi_{xx}(\mathbf{q},E^{(1)})(1-\widehat{q}_x^2) \right. \notag \\
&& + \left. \chi_{yy}(\mathbf{q},E^{(1)})(1-\widehat{q}_y^2) + \chi_{zz}(\mathbf{q},E^{(1)})(1-\widehat{q}_z^2) \right] , \notag \\\label{eqn3.1.2}
\end{eqnarray}
where the spin spectrum functions $\chi_{ii}(\mathbf{q},E)$ are given by
\begin{eqnarray}
\chi_{xx}(\mathbf{q},E) &=& \chi_{yy}(\mathbf{q},E) = \pi N S [\delta(E-\varepsilon_\mathbf{q}) - \delta(E+\varepsilon_{-\mathbf{q}}) ] , \notag \\
\chi_{zz}(\mathbf{q},E) &=& 2\pi \sum_{\mathbf{k}} [n_B(\varepsilon_{\mathbf{k}}) - n_B(\varepsilon_{\mathbf{k+q}})] \delta( E + \varepsilon_{\mathbf{k}} -\varepsilon_{\mathbf{k+q}}) . \notag \\
&&   \label{eqn3.1.3}
\end{eqnarray}
Here the transfer momentum and energy, $\mathbf{q}$ and $E^{(1)}$, are defined as in Eq. (\ref{eqn2.1.12}). While the transverse spin flips lead to single-magnon peak structures in the scattering probability, the longitudinal spin fluctuations contribute magnon density fluctuations. Besides these inelastic scattering contributions, there is one additional elastic scattering contribution from the magnon condensation, which gives 
\begin{equation}
\Gamma^{(1)}_{FM,c} = \frac{2\pi \vert g(\mathbf{q}) \vert^2 \Delta t }{\hbar} (N m_{FM})^2 \delta(E^{(1)})\delta_{\mathbf{q},0} (1-\widehat{q}_z^2) , \label{eqn3.1.3-2}
\end{equation}    
where $m_{FM}=\frac{1}{N}\sum_l \langle S^z_l \rangle$ is the ordered spin magnetic moment per site. It is noted that in experiment, $N$ is the number of local Wannier electron spins in the incident neutron beam. 
When considering only the single-magnon contributions without that of the magnon density fluctuations, the inelastic scattering probability of INS for the ordered ferromagnet follows
\begin{eqnarray}
\hspace*{-0.5cm} \Gamma^{(1)}_{FM} (\mathbf{q}_f,\mathbf{q}_i) &=& \frac{\pi N S \vert g(\mathbf{q}) \vert^2 \Delta t}{\hbar} n_B (E^{(1)}) (1+\widehat{q}_z^2) \notag \\
&& \times [\delta(E^{(1)}-\varepsilon_\mathbf{q}) - \delta(E^{(1)}+\varepsilon_{-\mathbf{q}}) ] . \label{eqn3.1.4}
\end{eqnarray}

Now let us study the coincidence probability of cINS for the ordered ferromagnet. Suppose the two incident thermal neutrons with momenta $\mathbf{q}_{i_1}$ and $\mathbf{q}_{i_2}$ are scattered into the states with momenta $\mathbf{q}_{f_1}$ and $\mathbf{q}_{f_2}$ and the incident neutrons are in spin states with $P_2 (\beta_{i_1},\beta_{i_2}) = P_1(\beta_{i_1}) P_1(\beta_{i_2})$ and $P_1(\beta_{i})$ defined in Eq. (\ref{eqn2.1.9}). Since the magnons are well-defined single-spin excitations with the momentum being a good quantum number, the coincidence probability of cINS for the ordered ferromagnet with only contributions from the single-magnon excitations has a product behavior described by Eq. (\ref{eqn3.1}), i.e.,  
\begin{eqnarray}
\Gamma^{(2)}_{FM} &=& \Gamma^{(1)}_{FM} (\mathbf{q}_{f_1},\mathbf{q}_{i_1}) \cdot \Gamma^{(1)}_{FM} (\mathbf{q}_{f_2},\mathbf{q}_{i_2}) \notag \\
&+& \Gamma^{(1)}_{FM} (\mathbf{q}_{f_1},\mathbf{q}_{i_2}) \cdot \Gamma^{(1)}_{FM} (\mathbf{q}_{f_2},\mathbf{q}_{i_1}), \label{eqn3.1.5}
\end{eqnarray}
where the four $\Gamma^{(1)}_{FM} (\mathbf{q}_{f},\mathbf{q}_{i})$'s are the scattering probabilities of the single-magnon relevant INS defined in Eq. (\ref{eqn3.1.4}). The magnon density fluctuations are not well-defined excitations, and their contribution would break down this simple product behavior.

\subsection{Antiferromagnet} \label{sec3.2}

Now let us consider an antiferromagnet in a cubic crystal lattice with long-range magnetic order. It has a spin lattice Hamiltonian defined by Eq. (\ref{eqn3.2}) with $J>0$. We introduce the spin rotation transformation as $\mathfrak{S}_l^x = e^{i \mathbf{Q}\cdot \mathbf{R}_l} S_l^x, \mathfrak{S}_l^y =  S_l^y , \mathfrak{S}_l^z = e^{i \mathbf{Q}\cdot \mathbf{R}_l} S_l^z $, where $\mathbf{Q}=(\pi/a,\pi/a, \pi/a)$ is the characteristic antiferromagnetic momentum. Introducing the Holstein-Primakoff transformation for the new spin operators, the spin Hamiltonian can be approximated in a linear spin-wave theory as 
\begin{equation}
H_{AF}={\sum_{\mathbf{k}}} ^{\prime}
\psi^\dag_\mathbf{k} 
\left( \begin{array}{cc}
A & B_\mathbf{k} \\
B_\mathbf{k} & A 
\end{array} \right) \psi_\mathbf{k} , 
\psi_\mathbf{k} = 
\left(\begin{array}{c}
a_\mathbf{k} \\
a^\dag_{-\mathbf{k}}
\end{array}
\right) , \label{eqn3.2.1}
\end{equation}  
where $A=JzS$, $B_\mathbf{k}=-JzS\gamma_\mathbf{k}$, and $\psi_\mathbf{k}$ is a bosonic Nambu spinor operator. Here the sum over $\mathbf{k}$ involves each pair $(\mathbf{k},-\mathbf{k})$ once. With the canonical transformation
\begin{equation}
\left(\begin{array}{c}
a_\mathbf{k} \\
a^\dag_{-\mathbf{k}}
\end{array}\right) = 
\left( \begin{array}{cc}
u_\mathbf{k} & v_\mathbf{k} \\
v_\mathbf{k} & u_\mathbf{k}
\end{array} \right)
\left(\begin{array}{c}
\beta_\mathbf{k} \\
\beta^\dag_{-\mathbf{k}}
\end{array}\right) , \label{eqn3.2.2}
\end{equation}  
the Hamiltonian can be diagonalized into the form
\begin{equation}
H_{AF} = {\sum_{\mathbf{k}}}^{\prime} E_\mathbf{k} ( \beta^\dag_\mathbf{k} \beta_\mathbf{k} + \beta^\dag_{-\mathbf{k}} \beta_{-\mathbf{k}} ),\label{eqn3.2.3} 
\end{equation}
where $E_\mathbf{k}= \sqrt{A^2-B_\mathbf{k}^2}$. Here $u^2_\mathbf{k}=\frac{A+E_\mathbf{k}}{2 E_\mathbf{k}}, v^2_\mathbf{k}=\frac{A-E_\mathbf{k}}{2 E_\mathbf{k}}$, and $u_\mathbf{k} v_\mathbf{k}=-\frac{B_\mathbf{k}}{2 E_\mathbf{k}}$.

It can easily be shown that the neutron scattering probability of INS for the ordered antiferromagnet $\Gamma^{(1)}_{AF} (\mathbf{q}_{f},\mathbf{q}_{i})$ follows an expression similar to Eq. (\ref{eqn3.1.2}) for $\Gamma^{(1)}_{FM} (\mathbf{q}_{f},\mathbf{q}_{i})$, with the corresponding spin spectrum functions $\chi_{ii}(\mathbf{q}, E)$ given by
\begin{eqnarray}
&& \chi_{xx}(\mathbf{q},E) = \chi_{yy}(\mathbf{q},E) \notag \\
=&& \pi N S \frac{A+B_\mathbf{q}}{E_\mathbf{q}}\left[ \delta(E - E_\mathbf{q} ) - \delta(E + E_\mathbf{q} ) \right]  \label{eqn3.2.4-1}   
\end{eqnarray}
and
\begin{widetext}
\begin{eqnarray}
\chi_{zz}(\mathbf{q},E) &=& 2\pi \sum_{\mathbf{k}} \left\{ [C^{(1)}_{\mathbf{k q}} \delta (E + \varepsilon_{\mathbf{k}} - \varepsilon_{\mathbf{k+q+Q}}) - C^{(4)}_{\mathbf{k q}} \delta (E - \varepsilon_{\mathbf{k}} + \varepsilon_{\mathbf{k+q+Q}})  ] [ n_B (\varepsilon_{\mathbf{k}}) - n_B(\varepsilon_{\mathbf{k+q+Q}})] \right. \notag \\
&& +  \left. [C^{(2)}_{\mathbf{k q}} \delta (E - \varepsilon_{\mathbf{k}} - \varepsilon_{\mathbf{k+q+Q}}) - C^{(3)}_{\mathbf{k q}} \delta (E + \varepsilon_{\mathbf{k}} + \varepsilon_{\mathbf{k+q+Q}})  ] [1 + n_B (\varepsilon_{\mathbf{k}}) + n_B(\varepsilon_{\mathbf{k+q+Q}})]  \right\} . \label{eqn3.2.4-2} 
\end{eqnarray}
Here $C^{(1)}_{\mathbf{k q}} = u^2_{\mathbf{k+q+Q}} u^2_{\mathbf{k}} + u_{\mathbf{k+q+Q}}v_{\mathbf{k+q+Q}} u_{\mathbf{k}} v_{\mathbf{k}}, C^{(2)}_{\mathbf{k q}} = u^2_{\mathbf{k+q+Q}} v^2_{\mathbf{k}} + u_{\mathbf{k+q+Q}}v_{\mathbf{k+q+Q}} u_{\mathbf{k}} v_{\mathbf{k}}, C^{(3)}_{\mathbf{k q}} = v^2_{\mathbf{k+q+Q}} u^2_{\mathbf{k}} + u_{\mathbf{k+q+Q}}v_{\mathbf{k+q+Q}} u_{\mathbf{k}} v_{\mathbf{k}}$, and $C^{(4)}_{\mathbf{k q}} = v^2_{\mathbf{k+q+Q}} v^2_{\mathbf{k}} + u_{\mathbf{k+q+Q}}v_{\mathbf{k+q+Q}} u_{\mathbf{k}} v_{\mathbf{k}}$. Similar to the ordered ferromagnet, there is also one additional elastic scattering contribution due to the magnon condensation, 
\begin{equation}
\Gamma^{(1)}_{AF,c} = \frac{2\pi \vert g(\mathbf{q}) \vert^2 \Delta t }{\hbar} (N m_{AF})^2 \delta(E^{(1)})\delta_{\mathbf{q},\mathbf{Q}} (1-\widehat{q}_z^2) , \label{eqn3.2.4-3}
\end{equation} 
where $m_{AF} = \frac{1}{N}\sum_l e^{i\mathbf{Q} \cdot \mathbf{R}_l} \langle S^z_l\rangle$ is the ordered antiferromagnetic moment per site. Here the transfer momentum and energy, $\mathbf{q}$ and $E^{(1)}$, are also defined in Eq. (\ref{eqn2.1.12}).
In the approximation with only the single-magnon contributions, the inelastic scattering probability of INS for the ordered antiferromagnet follows
\begin{equation}
\Gamma^{(1)}_{AF} (\mathbf{q}_{f},\mathbf{q}_{i}) = \frac{\pi N S \vert g(\mathbf{q}) \vert^2 \Delta t}{\hbar} \frac{A+B_\mathbf{q}}{E_\mathbf{q}} n_B(E^{(1)}) (1+\widehat{q}_z^2) [ \delta(E^{(1)} - E_\mathbf{q} ) - \delta(E^{(1)} + E_\mathbf{q} ) ]  . \label{eqn3.2.5} 
\end{equation}
\end{widetext} 

Now let us consider cINS with the thermal neutrons which have initial incident momenta $\mathbf{q}_{i_1}$ and $\mathbf{q}_{i_2}$ and final scattered momenta $\mathbf{q}_{f_1}$ and $\mathbf{q}_{f_2}$. The incident neutrons are independent, with the spin state defined by Eq. (\ref{eqn2.1.9}).  In linear spin-wave theory defined by the approximate Hamiltonian (\ref{eqn3.2.1}), the Nambu spinor operators with different momenta are decoupled. This means that the single-magnon excitations in the ordered antiferromagnet are decoupled. Therefore, in the linear spin-wave theory with only contributions from the single-magnon excitations, the conditions for the product behavior of the coincidence probability in Eq. (\ref{eqn3.1}) are also satisfied in the ordered antiferromagnet. In this approximation the coincidence probability of cINS for the ordered antiferromagnet follows a similar product behavior defined as  
\begin{eqnarray}
\Gamma^{(2)}_{AF} &=& \Gamma^{(1)}_{AF} (\mathbf{q}_{f_1},\mathbf{q}_{i_1}) \cdot \Gamma^{(1)}_{AF} (\mathbf{q}_{f_2},\mathbf{q}_{i_2}) \notag \\
&+& \Gamma^{(1)}_{AF} (\mathbf{q}_{f_1},\mathbf{q}_{i_2}) \cdot \Gamma^{(1)}_{AF} (\mathbf{q}_{f_2},\mathbf{q}_{i_1})  , \label{eqn3.2.6}
\end{eqnarray}
where the four $\Gamma^{(1)}_{AF} (\mathbf{q}_{f},\mathbf{q}_{i})$'s are the scattering probabilities of the single-magnon relevant INS defined in Eq. (\ref{eqn3.2.5}).

\section{Discussion and summary} \label{sec4}

In this paper we have proposed a coincidence technique, cINS, which has two neutron sources and two neutron detectors, with an additional coincidence detector. The two neutron sources emit two neutrons which are scattered by the electron spins of the magnetic material and are then detected by the two neutron detectors. The coincidence detector records the coincidence probability of the two scattered neutrons, which gives information on a two-spin Bethe-Salpeter wave function. This two-spin Bethe-Salpeter wave function defines the momentum-resolved dynamical wave function of the magnetic material with two spins excited. Thus, cINS can explicitly detect the two-spin magnetic correlations of the magnetic material. The coincidence probabilities of cINS for a ferromagnet and an antiferromagnet with long-range magnetic order have been calculated and show a product behavior contributed by the single-magnon relevant INSs. This trivial product behavior for the ordered ferromagnet and antiferromagnet is consistent with the magnetic properties dominated by the nearly free magnon excitations, which have no intrinsic two-spin magnetic correlations.

On the experimental instrument of cINS, we remark that the two incident neutrons can come from one neutron source. In this case the initial momenta of the two incident neutrons follow $\mathbf{q}_{i_2}=\mathbf{q}_{i_1} + \boldsymbol{\delta}_q$, with $\boldsymbol{\delta}_q \rightarrow 0$. These two incident neutrons can be regarded equivalently to be emitted from two different neutron sources but with nearly the same momenta. Thus, the theoretical formalism for cINS with one neutron source can be similarly established following the one we established in  Sec. \ref{sec2.2} for cINS with two neutron sources. There are two main challenges in the experimental realization of cINS. One is to develop a two-neutron coincidence detector, and the other one is accurate control of the coincidence detection. The two-photon coincidence measurement in modern quantum optics \citep{ShihPRL1995} and the coincidence detection of the photoelectron and the Auger electron in double-photoemission spectroscopy \citep{AliaevSS2018} may provide a useful guideline. 

The cINS we have proposed is one potential technique to study novel magnetic correlations which are far beyond the physics of the single-spin magnons. For example, the long-sought quantum spin liquids \citep{AndersonScience1987,PALeeRMP2006,BalentsNature2010,ZhouYiRMP2017} from strong frustration and quantum fluctuations show novel physics, such as various spin valence bond states \citep{XuBalentsPRB2011,ZhuWhitePRL2013,Ganesh2013PRB,Gong2013PRB} and novel quantum criticality \citep{SenthilScience2004}. Experimental study of the spin valence bond states by cINS would provide new insights into quantum spin liquids. The various quantum magnetic materials with spin dimers, such as  TlCuCl$_3$ \citep{CavadiniTlCuCl3-PRB2001}, SrCu$_2$(BO$_3$)$_2$ \citep{KageyamaSrCuBO-PRL1999}, and BaCuSi$_2$O$_6$ \citep{JaimeBaCuSiO-PRL2004}, could be the first focus in a cINS experiment. Quantum spin liquid materials in triangular, honeycomb, kagome, and hyperkagome lattices (e.g., the materials reviewed in Ref.  [\onlinecite{ZhouYiRMP2017,Chamorro2020}]) are also interesting target materials for a cINS experiment.

In summary, we have proposed a coincidence technique, cINS, which can explicitly detect the two-spin magnetic correlations of magnetic materials. It can be introduced to study the dynamical physics of the spin valence bond states of quantum magnets.

%%%%%%%%%%%%%%%%%%%%%%%%%%%%%%%%%%%%%%%%%%%%%%%%%%%%%%%%%
\section*{ACKNOWLEDGMENTS}
%{\it Acknowledgements}
We thank H. Shao and D. Z. Cao for invaluable discussions. This work was supported by the National Natural Science Foundation of China (Grants No. 11774299 and No. 11874318) and the Natural Science Foundation of Shandong Province (Grants No. ZR2017MA033 and No. ZR2018MA043).

%\newpage
\appendix

\section{Electron-neutron magnetic interaction}\label{secA1}

Let us review the electron-neutron magnetic interaction \citep{Lovesey1984,Squires1996,FelixPrice2013}. We define the neutron spin magnetic moment as $\boldsymbol{\mu}_n = -\gamma \mu_N \boldsymbol{\sigma}$, where $\gamma=1.91$ is a constant for the neutron gyromagnetic ratio, $\mu_N=\frac{e\hbar}{2 m_p}$ is the nuclear magneton, with $m_p$ being the proton mass, and $\boldsymbol{\sigma}$ is the Pauli matrix.  We define the electron spin magnetic moment as $\boldsymbol{\mu}_s = - g_s \mu_B \boldsymbol{S}$  and the electron orbital magnetic moment as $\boldsymbol{\mu}_l = - g_l \mu_B \boldsymbol{L}$, where the $g$ factors are set as $g_s = 2$ and $g_l = 1$ and $\mu_B = \frac{e\hbar}{2 m_e}$ is the Bohr magneton. The spin angular momentum operator $\mathbf{S}$ has eigenvalues $\pm \frac{1}{2}$, and the orbital angular momentum operator is defined as $\mathbf{L}=\frac{1}{\hbar}\mathbf{r}_e \times \mathbf{p}_e$. Suppose there is an electron at position $\mathbf{r}_e$ which can produce a magnetic field at position $\mathbf{r}_n$ as
\begin{equation}
\mathbf{B} = \frac{\mu_0}{4\pi} \mathbf{\nabla} \times \left[ (\boldsymbol{\mu}_s + \boldsymbol{\mu}_l) \times  \frac{\mathbf{R}}{R^3}\right] , \label{eqnA1.1}
\end{equation}  
where $\mu_0$ is the free-space permeability and $\mathbf{R}=\mathbf{r}_n - \mathbf{r}_e$. 
The electron-neutron magnetic interaction can be defined by $V = - \boldsymbol{\mu}_n \cdot \mathbf{B}$, which follows
\begin{equation}
V = - \frac{\mu_0}{4\pi} \gamma \mu_N \mu_B \boldsymbol{\sigma} \cdot \mathbf{\nabla} \times \left[ (g_s \boldsymbol{S} + g_l\boldsymbol{L}) \times  \frac{\mathbf{R}}{R^3} \right] . \label{eqnA1.2}
\end{equation}
Here we have introduced the orbital angular momentum $\mathbf{L}$ to describe the orbital motions of the electrons \citep{FelixPrice2013}. It is more convenient in the study of the orbital motions of electrons in compounds with transition metal and/or rare earth atoms.  

Let us present the second quantization of the electron-neutron magnetic interaction. Introduce the single-neutron states $\{\vert \mathbf{q}\beta\rangle\}$, where $\mathbf{q}$ is the neutron momentum and $\beta$ defines the neutron spin, and the single-electron states $\{\vert \lambda \rangle\}$, where $\lambda$ involves the momentum, orbital, and spin degrees of freedom, etc. Let us introduce the following identities:
\begin{equation*}
1=\frac{1}{V_1} \int d\mathbf{r}_e \vert \mathbf{r}_e \rangle\langle \mathbf{r}_e \vert 
\end{equation*}
for the electrons, and 
\begin{equation*}
1=\frac{1}{V_2} \int d\mathbf{r}_n \vert \mathbf{r}_n \rangle\langle \mathbf{r}_n \vert 
\end{equation*}
for the neutrons. Here $V_{1}$ and $V_{2}$ are the renormalization volumes for the single-electron and single-neutron states, respectively. The electron-neutron magnetic interaction in second quantization can be expressed as
\begin{equation}
\widehat{V} = \widehat{V}_s + \widehat{V}_l , \label{eqnA1.3} 
\end{equation}
where 
\begin{eqnarray}
&&\widehat{V}_s = \frac{4\pi A_s}{V_2}\sum_{\mathbf{q}_i \mathbf{q}_f} \widehat{\boldsymbol{\sigma}}_{\mathbf{q}_f \mathbf{q}_i} \cdot [ \widehat{\mathbf{q}}\times (\mathbf{D}^s(\mathbf{q})\times \widehat{\mathbf{q}}) ] , \label{eqnA1.4.1}  \\
&&\widehat{V}_l = \frac{4\pi A_l}{V_2}\sum_{\mathbf{q}_i \mathbf{q}_f} \widehat{\boldsymbol{\sigma}}_{\mathbf{q}_f \mathbf{q}_i} \cdot [ \widehat{\mathbf{q}}\times (\mathbf{D}^l(\mathbf{q})\times \widehat{\mathbf{q}}) ] . \label{eqnA1.4.2} 
\end{eqnarray}
Here the momentum $\mathbf{q} = \mathbf{q}_f - \mathbf{q}_i$, and $\widehat{\mathbf{q}} = \frac{\mathbf{q}}{q}$. 
It is noted that $\widehat{\mathbf{q}}\times (\mathbf{D}(\mathbf{q})\times \widehat{\mathbf{q}})$ can be reexpressed as $\mathbf{D}_{\perp}(\mathbf{q})$:     
\begin{equation}
\mathbf{D}_{\perp}(\mathbf{q}) =  \mathbf{D}(\mathbf{q}) - \widehat{\mathbf{q}} (\mathbf{D}(\mathbf{q})\cdot \widehat{\mathbf{q}}) . \label{eqnA1.5}
\end{equation} 
In the electron-neutron magnetic interaction $\widehat{V}$, the constants $A_{s}$ and $A_{l}$ are defined as
\begin{equation}
 A_s = - \frac{\mu_0}{4\pi}\gamma g_s \mu_N \mu_B, A_l = - \frac{\mu_0}{4\pi}\gamma g_l \mu_N \mu_B , \label{eqnA1.6}
\end{equation}
and the operator $\widehat{\boldsymbol{\sigma}}_{\mathbf{q}_f \mathbf{q}_i}$ is defined as
\begin{equation}
\widehat{\boldsymbol{\sigma}}_{\mathbf{q}_f \mathbf{q}_i} = \sum_{\beta_i \beta_f} d^{\dag}_{\mathbf{q}_f \beta_f} \boldsymbol{\sigma}_{\beta_f \beta_i} d_{\mathbf{q}_i \beta_i} , \label{eqnA1.7} 
\end{equation}
where $d_{\mathbf{q} \beta}$ and $d^{\dag}_{\mathbf{q} \beta}$ are the annihilation and creation operators for the neutrons. The operators $\mathbf{D}^{s}$ and $\mathbf{D}^{l}$ in $\widehat{V}$ are defined as
\begin{eqnarray}
&&\mathbf{D}^s(\mathbf{q}) = \sum_{\lambda_1 \lambda_2} c_{\lambda_2}^{\dag} \mathbf{M}^{(s)}_{\lambda_2 \lambda_1}(\mathbf{q}) c_{\lambda_1} , \label{eqnA1.8.1} \\
&&\mathbf{D}^l (\mathbf{q})= \sum_{\lambda_1 \lambda_2} c_{\lambda_2}^{\dag} \mathbf{M}^{(l)}_{\lambda_2 \lambda_1}(\mathbf{q}) c_{\lambda_1} ,  \label{eqnA1.8.2}
\end{eqnarray}
where $c_{\lambda}$ and $c^\dag_{\lambda}$ are the annihilation and creation operators for the electrons and 
\begin{eqnarray*}
\mathbf{M}^{(s)}_{\lambda_2 \lambda_1}(\mathbf{q}) = \frac{1}{V_1}\int d\mathbf{r}_e [\psi^{\ast}_{\lambda_2}(\mathbf{r}_e) \mathbf{S} \psi_{\lambda_1} (\mathbf{r}_e)] e^{-i\mathbf{q}\cdot \mathbf{r}_e} , \\ %\label{eqnA1.9.1} \\
\mathbf{M}^{(l)}_{\lambda_2 \lambda_1}(\mathbf{q}) = \frac{1}{V_1}\int d\mathbf{r}_e [\psi^{\ast}_{\lambda_2}(\mathbf{r}_e) \mathbf{L} \psi_{\lambda_1} (\mathbf{r}_e)] e^{-i\mathbf{q}\cdot \mathbf{r}_e} .  %\label{eqnA1.9.2}
\end{eqnarray*}
Here $\psi_{\lambda} (\mathbf{r}_e)$ is the single-electron wave function. 
 
Let us focus on the spin degrees of freedom of the electrons and ignore the orbital ones. We consider the electrons to be in the local Wannier states $\{\vert l \alpha \rangle\}$ with position $\mathbf{R}_{l}$ and spin $\alpha$. $\mathbf{D}^{s}(\mathbf{q})$ can be approximately defined as 
\begin{equation}
\mathbf{D}^{s}(\mathbf{q}) = F_0(\mathbf{q}) \mathbf{S}(\mathbf{q}) ,     \label{eqnA1.10}
\end{equation} 
where the spin operator $\mathbf{S}(\mathbf{q})$ is defined as
\begin{equation}
\mathbf{S}(\mathbf{q}) = \sum_{l} \mathbf{S}_l e^{-i\mathbf{q}\cdot \mathbf{R}_l}, \mathbf{S}_l =\sum_{\alpha_1 \alpha_2} c^{\dag}_{l\alpha_2} \mathbf{S}_{\alpha_2 \alpha_1} c_{l\alpha_1} , \label{eqnA1.11}
\end{equation}
and the magnetic form factor $F_0(\mathbf{q})$ is given by
\begin{equation}
F_0(\mathbf{q}) = \frac{1}{V_1}\int d\mathbf{a} \psi^{\ast}_{l}(\mathbf{a}) \psi_{l}(\mathbf{a}) e^{-i\mathbf{q}\cdot \mathbf{a}}, \mathbf{a} = \mathbf{r}_e - \mathbf{R}_l . \label{eqnA1.12} 
\end{equation}
Here we have made an approximation to consider only the on-site intraorbital integrals and ignore all the other contributions. 
For the itinerant electrons in the Bloch states $\{\vert \mathbf{k} \alpha \rangle\}$, the operator $\mathbf{D}^{s}(\mathbf{q})$ can be given by 
\begin{equation}
\mathbf{D}^{s}(\mathbf{q}) = \sum_{\mathbf{k}_1 \mathbf{k}_2} F_{\mathbf{k}_2 \mathbf{k}_1}(\mathbf{q}) \mathbf{S}_{\mathbf{k}_2 \mathbf{k}_1} ,  \label{eqnA1.13}  
\end{equation}
where the spin operator is defined by 
\begin{equation}
\mathbf{S}_{\mathbf{k}_2 \mathbf{k}_1} = \sum_{\alpha_1 \alpha_2} c^{\dag}_{\mathbf{k}_2\alpha_2} \mathbf{S}_{\alpha_2 \alpha_1} c_{\mathbf{k}_1\alpha_1} , \label{eqnA1.14}
\end{equation}
and the form factor $F_{\mathbf{k}_2 \mathbf{k}_1}(\mathbf{q})$ is given by
\begin{equation}
F_{\mathbf{k}_2 \mathbf{k}_1}(\mathbf{q}) = \frac{1}{V_1}\int d\mathbf{r}_e \psi^{\ast}_{\mathbf{k}_2}(\mathbf{r}_e) \psi_{\mathbf{k}_1}(\mathbf{r}_e) e^{-i\mathbf{q}\cdot\mathbf{r}_e} . \label{eqnA1.15}
\end{equation}
Here $\psi_{\mathbf{k}}(\mathbf{r}_e)$ is the Bloch-state wave function. In the approximation with $\psi_{\mathbf{k}}(\mathbf{r}_e) = e^{i\mathbf{k}\cdot \mathbf{r}_e}$, $\mathbf{D}^{s}(\mathbf{q})$ can be simplified as 
\begin{equation}
\mathbf{D}^{s}(\mathbf{q}) = \sum_{\mathbf{k}} \mathbf{S}_{\mathbf{k},\mathbf{k+q}} .  \label{eqnA1.16}  
\end{equation}

In summary, the electron-neutron magnetic interaction with only the spin degrees of freedom of the electrons can be given as follows. 
For the local Wannier electrons, 
\begin{equation}
\widehat{V}_s = \sum_{\mathbf{q}_i \mathbf{q}_f} g(\mathbf{q}) \widehat{\boldsymbol{\sigma}}_{\mathbf{q}_f \mathbf{q}_i} \cdot \mathbf{S}_{\perp}(\mathbf{q}) ,   \label{eqnA1.17} 
\end{equation}
where $g(\mathbf{q})\equiv g F_0(\mathbf{q})$, with $g  = \frac{4\pi A_s}{V_2}$, and $\mathbf{S}_{\perp}(\mathbf{q})$ is the projection of $\mathbf{S}(\mathbf{q})$ in the perpendicular plane normal to the momentum $\mathbf{q}$ and is defined similarly to $\mathbf{D}_{\perp}(\mathbf{q})$ in Eq. (\ref{eqnA1.5}).
For the itinerant Bloch electrons, 
\begin{equation}
\widehat{V}_s = \sum_{\mathbf{q}_i \mathbf{q}_f \mathbf{k}_1 \mathbf{k}_2} g_{\mathbf{k}_2 \mathbf{k}_1}(\mathbf{q}) \widehat{\boldsymbol{\sigma}}_{\mathbf{q}_f \mathbf{q}_i} \cdot \mathbf{S}_{\mathbf{k}_2 \mathbf{k}_1,\perp} ,    \label{eqnA1.18} 
\end{equation}
where $g_{\mathbf{k}_2 \mathbf{k}_1}(\mathbf{q})\equiv g F_{\mathbf{k}_2 \mathbf{k}_1}(\mathbf{q})$ and $ \mathbf{S}_{\mathbf{k}_2 \mathbf{k}_1,\perp}$ is defined similarly to $\mathbf{D}_{\perp}(\mathbf{q})$ in Eq. (\ref{eqnA1.5}).
It should be noted that the form factors $F_0(\mathbf{q})$ and $F_{\mathbf{k}_2 \mathbf{k}_1}(\mathbf{q})$ have strong $\mathbf{q}$ dependence. 
 
One remark is that in the above electron-neutron magnetic interaction, the contributions from the spin and orbital magnetic moments are independently derived. 
In this case, the spin-orbit coupling is weak like for the electrons of the transition metal atoms. In the case with strong spin-orbit coupling such as that of the electrons of rare earth atoms, the total angular momentum $\mathbf{J}$ is conserved. In this case we can introduce the total magnetic moment $\boldsymbol{\mu}_J = -g(JLS) \mu_B \mathbf{J}$, with the Land\'{e} $g$ factor $g(JLS)$ defined following $g_l \mathbf{L} + g_s \mathbf{S} = g(JLS) \mathbf{J}$. A similar derivation can give us an electron-neutron magnetic interaction in this case. Another remark is that the Debye-Waller factor \citep{Lovesey1984,Squires1996} from the crystal lattice effects is ignored in our discussion on the neutron scattering probability of the inelastic neutron scattering.  

\section{Calculations for scattering probability of INS} \label{secA2}

Let us introduce the imaginary-time Green's functions $G_{ij}(\mathbf{q},\tau) = -\langle T_\tau	S_i (\mathbf{q},\tau) S^{\dag}_j (\mathbf{q},0) \rangle $ with $i,j=x,y,z$. The corresponding spectrum functions are defined as $\chi_{ij} (\mathbf{q},E) = - 2 \text{ Im} G_{ij}(\mathbf{q},i\nu_n \rightarrow E + i\delta^+)$. Then we have 
\begin{equation}
G(\mathbf{q},\tau) = \sum_{ij} G_{ij}(\mathbf{q},\tau) (\delta_{ij} - \widehat{q}_i \widehat{q}_j ) , \label{eqnA2.1} 
\end{equation}
and 
\begin{equation}
\chi (\mathbf{q},E) = \sum_{ij} \chi_{ij} (\mathbf{q},E) (\delta_{ij} - \widehat{q}_i \widehat{q}_j ) . \label{eqnA2.2} 
\end{equation}

First, let us consider the ferromagnet in a cubic crystal lattice with a long-range magnetic order. We introduce the imaginary-time Green's function for the ferromagnetic magnons, $G_a(\mathbf{q},\tau) = -\langle T_\tau a_\mathbf{k}(\tau) a^{\dag}_\mathbf{k} (0) \rangle$. Its frequency Fourier transformation is given by
\begin{equation}
G_a(\mathbf{k},i\nu_n) = \frac{1}{i\nu_n - \varepsilon_\mathbf{k}} , \label{eqnA2.3}
\end{equation}
where the magnon energy dispersion $\varepsilon_\mathbf{k}$ is defined in Eq. (\ref{eqn3.1.1}). It can be shown that in the linear spin-wave approximation,  
\begin{eqnarray}
&& G_{xx}(\mathbf{q},i\nu_n) = G_{yy}(\mathbf{q},i\nu_n) \nonumber \\
&=& \frac{NS}{2}\left[ G_a(\mathbf{q},i\nu_n) + G_a(-\mathbf{q},-i\nu_n)\right] \label{eqnA2.4}
\end{eqnarray}
and 
\begin{eqnarray}
G_{zz}(\mathbf{q},i\nu_n) &=& -\frac{1}{\beta} \sum_{\mathbf{k}, i\nu_1} G_a(\mathbf{k}+\mathbf{q},i\nu_1 + i\nu_n) G_a(\mathbf{k},i\nu_1) . \nonumber \\
&&  \label{eqnA2.5} 
\end{eqnarray}
The other Green's functions follow
\begin{equation}
G_{ij}(\mathbf{q},i\nu_n) = 0, \text{ for } i \not= j . \label{eqnA2.6} 
\end{equation}
From these results, we can obtain the spectrum functions $\chi_{ij}(\mathbf{q},E)$ in Eq. (\ref{eqn3.1.3}) for the ordered ferromagnet.

Now let us consider the antiferromagnet in a cubic crystal lattice with a long-range magnetic order. We introduce the imaginary-time Green's function of a Nambu spinor operator, 
\begin{equation}
G_\psi (\mathbf{k},\tau) = -\langle T_\tau \psi_\mathbf{k}(\tau) \psi^{\dag}_\mathbf{k}(0)\rangle, \label{eqnA2.7}
\end{equation}
where $\psi_\mathbf{k}$ is defined in Eq. (\ref{eqn3.2.1}). It can be shown that the frequency Green's function follows
\begin{equation}
G_\psi (\mathbf{k},i\nu_n) = \frac{i\nu_n \tau_3 + A - B_\mathbf{k} \tau_1}{(i\nu_n)^2 - E^{2}_\mathbf{k}} , \label{eqnA2.8}
\end{equation}
where $A$ and $B_\mathbf{k}$ are defined in Eq. (\ref{eqn3.2.1}) and the magnon energy $E_\mathbf{k}$ is given in Eq. (\ref{eqn3.2.3}). Here $\tau_i \text{ } (i=1,2,3)$ are the Pauli matrices. 

It can be shown that 
%\begin{widetext}
\begin{eqnarray}
G_{xx} (\mathbf{q},i\nu_n) &=& \frac{N S}{2} \mathrm{Tr}[ G_{\psi} (\mathbf{q}+\mathbf{Q},i\nu_n) +  G_{\psi} (\mathbf{q}+\mathbf{Q},i\nu_n) \tau_1 ] , \nonumber \\
G_{yy} (\mathbf{q},i\nu_n) &=& \frac{N S}{2} \mathrm{Tr}[ G_{\psi} (\mathbf{q},i\nu_n) - G_{\psi} (\mathbf{q},i\nu_n) \tau_1 ] , \label{eqnA2.9} 
\end{eqnarray}
and 
\begin{widetext}
\begin{equation}
G_{zz} (\mathbf{q},i\nu_n) = -\frac{1}{\beta}\sum_{\mathbf{k}, i\nu_1} [G^{(11)}_\psi (\mathbf{k}+\mathbf{q}+\mathbf{Q},i\nu_1 + i\nu_n) G^{(11)}_\psi (\mathbf{k},i\nu_1) + G^{(21)}_\psi (\mathbf{k}+\mathbf{q}+\mathbf{Q},i\nu_1 + i\nu_n) G^{(12)}_\psi (\mathbf{k},i\nu_1)] . \label{eqnA2.10}    
\end{equation}
\end{widetext}
The other Green's functions $G_{ij}(\mathbf{q},i\nu_n) = 0$ for the cases with $i \not= j$. With these results, we can obtain the spectrum functions $\chi_{ij}(\mathbf{q}, E)$ in Eqs. (\ref{eqn3.2.4-1}) and (\ref{eqn3.2.4-2}) for the ordered antiferromagnet.

%%%%%%%%%%%%%%%%%%%%%%%%%%%%%%%%%%%%%%%%%%%%%%%%%%%%%%%%%%%%%%%%%%%%%%%%%%%%%%%%%%

%\bibliography{references}

\end{document}